\input harvmac

\def\MM{{\cal M}}
\def\tH{{\tilde H}}

\def\journal#1&#2(#3){\unskip, \sl #1\ \bf #2 \rm(19#3) }
\def\andjournal#1&#2(#3){\sl #1~\bf #2 \rm (19#3) }

\def\ie{{\it i.e.}}
\def\eg{{\it e.g.}}

\def\tilde{\widetilde}

\def\frac#1#2{{#1\over#2}}

\def\inbar{\,\vrule height1.5ex width.4pt depth0pt}
\def\IC{\relax\hbox{$\inbar\kern-.3em{\rm C}$}}
\def\IR{\relax{\rm I\kern-.18em R}}
\def\IP{\relax{\rm I\kern-.18em P}}

%
%
\def\np#1#2#3{Nucl. Phys. {\bf B#1} (#2) #3}
\def\pl#1#2#3{Phys. Lett. {\bf #1B} (#2) #3}

\def\prl#1#2#3{Phys. Rev. Lett. {\bf #1} (#2) #3}

\def\prd#1#2#3{Phys. Rev. {\bf D#1} (#2) #3}

\def\cqg#1#2#3{Class. Quant. Grav. {\bf #1} (#2) #3}

\def\atmp#1#2#3{Adv. Theor. Math. Phys. {\bf #1} (#2) #3}
\def\jhep#1#2#3{J. High Energy Phys. {\bf #1} (#2) #3}

\catcode`\@=11
\def\slash#1{\mathord{\mathpalette\c@ncel{#1}}}
\overfullrule=0pt

\def\NN{{\cal N}}

\def\WW{{\cal W}}

\def\underrel#1\over#2{\mathrel{\mathop{\kern\z@#1}\limits_{#2}}}

\catcode`\@=12


%


\lref\wittenstrings{E. Witten, talk given at Strings '98, {\tt 
http://www.itp.ucsb.edu/online/strings98/ witten}.}
\lref\juan{J. M. Maldacena, ``The Large $N$ Limit of Superconformal Field 
Theories and Supergravity'', hep-th/9711200.}
\lref\natistr{N. Seiberg, ``New Theories in Six Dimensions and 
Matrix Description of M-Theory on $T^5$ and $T^5/Z_2$'', hep-th/9705221,
\pl{408}{1997}{98}.}
\lref\brs{M. Berkooz, M. Rozali and N. Seiberg, ``Matrix Description of 
M Theory on $T^4$ and $T^5$'', hep-th/9704089, \pl{408}{1997}{105}.}
\lref\oens{O. Aharony, M. Berkooz, S. Kachru, N. Seiberg, E. Silverstein,
``Matrix Description of Interacting Theories in Six Dimensions'',
hep-th/9707079, \atmp{1}{1998}{148}.}
\lref\edhiggs{E. Witten, ``On the Conformal Field Theory of the Higgs 
Branch'', hep-th/9707093, \jhep{7}{1997}{3}.}
\lref\andyjuan{J.M. Maldacena and A. Strominger, ``Semiclassical Decay 
of Near Extremal Five-Branes'', hep-th/9710014, \jhep{12}{1997}{8}.}
\lref\chs{C. Callan, J. Harvey and A. Strominger, ``Supersymmetric
String Solitons'', hep-th/9112030,
in Trieste 1991, Proceedings, String Theory and Quantum Gravity 1991, 
208.}
\lref\cghs{C. Callan, S.B. Giddings, J.A. Harvey and A. Strominger,
``Evanescent Black Holes'', hep-th/9111056, \prd{45}{1992}{1005}.}
\lref\sunny{N. Itzhaki, J.M. Maldacena, J. Sonnenschein and 
S. Yankielowicz, ``Supergravity and the Large $N$ Limit of Theories
with Sixteen Supercharges'', hep-th/9802042, \prd{48}{1998}{46}.
}
\lref\townsend{H.J. Boonstra, K. Skenderis and P.K. Townsend, ``The 
domain-wall/QFT Correspondence'', hep-th/9807137.}
\lref\gkp{S.S. Gubser, I.R. Klebanov and A.M. Polyakov, ``Gauge 
Correlators from Noncritical String Theory'', hep-th/9802109.}
\lref\wttn{E. Witten, ``Anti-de-Sitter Space and Holography'', 
hep-th/9802150, \pl{428}{1998}{105}.}
\lref\sushol{L. Susskind, ``The World as a Hologram'', hep-th/9409089,
J. Math. Phys. {\bf 36} (1995) 6377.}
\lref\tfthol{C.T. Stephens, G. 't Hooft and B.F. Whiting, ``Black Hole 
Evaporation Without Information Loss'', gr-qc/9310006,
\cqg{11}{1994}{621}.}
\lref\thorn{C. Thorn, ``Reformulating String Theory with the $1/N$
Expansion'', 
hep-th/9405069, published in
Sakharov Conf. on Physics, Moscow (1991) 447.}
\lref\orthooft{G. 't Hooft, ``Dimensional Reduction in Quantum
Gravity'', gr-qc/9310026, published in Salamfest 1993, 284.}
\lref\polams{A.M. Polyakov, ``String Theory and Quark Confinement'',
hep-th/9711002, Nucl. Phys. Proc. Suppl. {\bf 68} (1998) 1.}
\lref\witsus{L. Susskind and E. Witten, ``The Holographic Bound in 
Anti-De-Sitter Space'', hep-th/9805114.}
\lref\morgins{P. Ginsparg and G. Moore, ``Lectures on 2-D Gravity
and 2-D String Theory'', hep-th/9304011, TASI 92 summer school, Published
in TASI 92:277-470.}
\lref\abs{O. Aharony, M. Berkooz and N. Seiberg, ``Light Cone
Description of $(2,0)$ Superconformal Theories in Six Dimensions'',
hep-th/9712117,
\atmp{2}{1998}{119}.}
\lref\wittensix{E. Witten, ``New `Gauge' Theories in Six Dimensions'',
hep-th/9710065, \jhep{1}{1998}{1}, \atmp{2}{1998}{61}.}
\lref\aoy{O. Aharony, Y. Oz and Z. Yin, ``M Theory on $AdS_p \times
S^{11-p}$ and Superconformal Field Theories'', hep-th/9803051,
\pl{430}{1998}{87}.}
\lref\lr{R. G. Leigh and M. Rozali, ``The Large $N$ Limit of the
$(2,0)$ Superconformal Field Theory'', hep-th/9803068, 
\pl{431}{1998}{311}.}
\lref\minwalla{S. Minwalla, ``Particles on $AdS_{4/7}$ and Primary
Operators on M2-brane and M5-brane World Volumes'', hep-th/9803053.}
\lref\halyo{E. Halyo, ``Supergravity on $AdS_{4/7} \times S^{7/4}$ and
M Branes'', hep-th/9803077, \jhep{4}{1998}{11}.}
\lref\edbaryon{E. Witten, ``Branes and Baryons in Anti-De Sitter Space'',
hep-th/9805112.}
\lref\gh{S. S. Gubser and A. Hashimoto, ``Exact Absorption Probabilities
for the D3-brane'', hep-th/9805140.}
\lref\ks{D. Kutasov and N. Seiberg, ``Non-Critical Superstrings'', 
\pl{251}{1990}{67}.}
\lref\wttnmqcd{E. Witten, ``Solutions of Four-Dimensional Field Theories 
Via M-Theory'', hep-th/9703166,
\np{500}{1997}{3}.}
\lref\wttnbh{E. Witten, ``On String Theory and Black Holes'', 
\prd{44}{1991}{314}.}
\lref\eliezer{S. Elitzur, A. Forge and E. Rabinovici, ``Some Global
Aspects of String Compactification'', \np{359}{1990}{344}.}
\lref\natinotes{N. Seiberg, ``Notes on Quantum Liouville Theory and 
Quantum Gravity'', Prog. Theor. Phys. Suppl. {\bf 102} (1990) 319.}
\lref\joenotes{J. Polchinski, ``Remarks on the Liouville Field Theory'',
Strings '90 conf., Published in Coll. Station Workshop, 1990.}
\lref\gb{T. Banks and M.B. Green, ``Non-Perturbative Effects in AdS in 
Five-Dimensions $\times S^5$ String Theory and D=4 SUSY Yang-Mills'',
hep-th/9804170, \jhep{05}{1998}{2}.}
\lref\dh{A. Dabholkar, J.A. Harvey, ``Nonrenormalization of the 
Superstring Tension", \prl{63}{1989}{478}.}
\lref\diaco{D.-E. Diaconescu and N. Seiberg, ``The Coulomb Branch of 
(4,4) Supersymmetric Field Theories in Two Dimensions'',
hep-th/9707158, \jhep{07}{1997}{1}.}
\lref\bklt{V. Balasubramanian, P. Kraus, A. Lawrence, S. P. Trivedi,
``Holographic Probes of Anti-De Sitter Spacetimes'', hep-th/9808017.}
\lref\wittenmqcdone{E. Witten, ``Branes and the Dynamics of QCD'', 
hep-th/9706109, \np{507}{1997}{658}.}
\lref\juanwilson{J. Maldacena, ``Wilson Loops in Large $N$ Field
Theories'', hep-th/9803002, \prl{80}{1998}{4859}.}
\lref\bgkr{M. Bianchi, M.B. Green, S. Kovacs and G. Rossi, ``Instantons
in Supersymmetric Yang-Mills and D-instantons in IIB Superstring
Theory'', hep-th/9807033.}


%
\rightline{EFI-98-39, RU-98-38, IASSNS-HEP-98-75}
\Title{
\rightline{hep-th/9808149}
} {\vbox{\centerline{Linear Dilatons, NS5-branes and Holography}}}
\medskip

\centerline{\it
Ofer Aharony${}^1$, Micha Berkooz${}^2$, David Kutasov${}^3$,
and Nathan Seiberg${}^2$}
\bigskip
\centerline{${}^1$Department of Physics and Astronomy, 
Rutgers University, Piscataway, NJ 08855}
\centerline{${}^2$School of Natural Sciences,
Institute for Advanced Study, Olden Lane, Princeton, NJ 08540}
\centerline{${}^3$Department of Physics, University of Chicago,
5640 S. Ellis Av., Chicago, IL 60637}

\smallskip

\vglue .3cm
\bigskip

\noindent
We argue that vacua of string theory which asymptote at weak coupling
to linear dilaton backgrounds are holographic.  The full string theory
in such vacua is ``dual'' to a theory without gravity in fewer
dimensions. The dual theory is generically not a local quantum field
theory.  Excitations of the string vacuum, which can be studied in the
weak coupling region using worldsheet methods, give rise to
observables in the dual theory.  An interesting example is string
theory in the near-horizon background of parallel NS5-branes, the CHS
model, which is dual to the decoupled NS5-brane theory (``little
string theory'').  This duality can be used to study some of the
observables in this theory and some of their correlation functions.
Another interesting example is the ``old'' matrix model, which gives
a holographic description of two dimensional string theory.

\Date{8/98}

\newsec{Introduction and Discussion}

It has been suggested by 't Hooft \orthooft\ and Susskind \sushol\
(see also \thorn) that
any consistent quantum theory of gravity must be holographic, \ie\
the number of degrees of freedom in any spatial domain is finite,
proportional to the area of the boundary of the domain
in Planck units.  This is unlike standard local quantum field
theories in two respects.  First, the number of degrees of freedom of
quantum field theory is proportional to the volume of the system
rather than the area of its boundary.  
Second, continuum quantum field theories have
an infinite number of degrees of freedom per unit volume.  The latter
difference is presumably responsible for the finiteness of string
theory, and appears like a built-in cutoff.  The first is more
surprising.  It basically states that the degrees of freedom in a
certain region ``live'' on the boundary of the region rather than in
the interior.  Equivalently, in any generally covariant theory it is
difficult to define local observables, and therefore it is natural to
assume that there are no such observables.  The only observables
should exist on the boundary.

In some vague sense this reduction by one dimension is quite familiar
in string theory.  It is widely believed that string theory, as a
theory which describes all particles and interactions, has only
on-shell information.  The theory cannot be probed by sources which
are not within the theory itself, and hence we cannot probe it
off-shell.  Therefore, in string theory we usually compute S-matrix
elements rather than Green functions.  In $d+1$ space-time dimensions
the on-shell momenta have only $d$ independent components while the
off-shell momenta have $d+1$ components.  This suggests that an
on-shell theory like string theory in $d+1$ dimensions can be equivalent 
to an off-shell theory in $d$ dimensions.  In particular, Polyakov
suggested that a field theory like QCD in $d$ dimensions can be
equivalent to a $d+1$ dimensional string theory \polams.

Recently these ideas have been made much more concrete. String theory
and M-theory in anti-de-Sitter space \juan\ beautifully demonstrate
holography \witsus. The theory on the boundary is in
this case a local quantum field theory, whose observables are
correlation functions of local operators.  In the bulk theory they
describe the response of the theory to disturbances at infinity 
\refs{\gkp,\wttn}.

The purpose of this note is to study a class of string backgrounds
which exhibit holography but whose boundary dynamics is in general
not local.
Specifically, we will discuss linear dilaton backgrounds which
asymptote, as a space-like coordinate $\phi
\rightarrow \infty$, to spacetimes of the form
$\MM\times \IR^{d,1}$ where 
$\MM$ is a compact space and $\phi$ is one of the coordinates in
$\IR^{d,1}$. The string frame metric and string coupling
asymptote to
\eqn\strngm{\eqalign{ds^2_{string}=&dx^2+d\phi^2+ds^2(\MM)\cr
g_s^2=&e^{-Q\phi}\cr}}
where $x $ is a coordinate on $\IR^{d-1,1}$ and the string metric on
$\MM$ is independent of $x$ and $\phi$.  We will also comment on the
case where some of the coordinates of $\IR^{d-1,1}$ are compactified
on a torus.  

We propose that any string background that behaves
asymptotically as \strngm\ is equivalent to 
a lower dimensional off-shell
theory without gravity whose observables 
live at the boundary $\phi\to\infty$. Off-shell physical observables
in the ``boundary'' theory are identified with on-shell physical
excitations in the string background \strngm. Note that 
this proposal is in agreement with
a known property of string vacua which asymptote to
\strngm\ \natinotes: the profiles of physical string excitations,
\eg\ those described on the worldsheet 
by BRST invariant vertex operators, are non-normalizable
and supported (typically exponentially in $\phi$) at $\phi\to\infty$.  

Green functions in the off-shell boundary theory 
are identified, as in the AdS/CFT correspondence, 
with on-shell amplitudes in string theory.  Perturbatively they are
given by worldsheet correlation functions of the corresponding
physical vertex operators. It is well known that generic
worldsheet correlation functions are sensitive to the  
spacetime background at finite $\phi$ and not just to the
asymptotic form \strngm. On the worldsheet this is the statement
that higher genus contributions to correlation functions
become more and more important as $\phi\to-\infty$. In spacetime
this can be seen by analyzing the metric \strngm\ in the Einstein frame,
\eqn\strngme{ds^2_{Einstein}=e^{\beta \phi}(dx^2+d\phi^2+ds^2(\MM)),}
where the positive number $\beta$ depends on $Q$, $d$ and the
dimension of $\MM$.  As $\phi \rightarrow \infty$ the distances between
points at fixed $x$ in $\IR^{d-1,1}$ diverge, as in the AdS examples.
Therefore, disturbances on the boundary have to propagate to the bulk
before they can interact.  This is a necessary condition for
holography \wittenstrings.  Equivalently, in the string frame the
distances remain finite but the string interactions vanish as
$\phi \rightarrow \infty$.  
Again, signals have to propagate to the bulk in order to
interact.

To summarize, the description of 
the bulk theory \strngm\ is useful near $\phi 
\approx \infty$, where the string theory is weakly coupled and
excitations, such as BRST invariant vertex operators and D-branes, 
can be studied using worldsheet methods.
Holography relates these excitations to observables in the boundary
theory. To compute correlation functions some information about
the strong coupling region $\phi\to-\infty$ is needed. 

There are several known classes of string vacua which asymptote
to \strngm. Different vacua utilize different mechanisms for
regulating the divergences in the strong coupling region.
In Liouville theory (for reviews see 
\refs{\natinotes,\joenotes,\morgins}) a tachyon condensate, a
potential on the worldsheet, makes it harder for the strings to
propagate to the strong coupling region.  In the two dimensional black
hole \refs{\eliezer,\wttnbh}, which is the $SL(2,\IR)/U(1)$ coset
theory, the spacetime topology at finite $\phi$ is modified and the
string coupling is bounded.  Finally, as we will see below, in the
theory of NS5-branes the resolution of the strong coupling
singularity cannot be understood using worldsheet methods.
String duality can be used to show that in some cases the low
energy theory becomes weakly coupled in other variables.

When some of the coordinates $x$ in \strngm\ are compactified on a
torus, the underlying string theory enjoys a T-duality symmetry, implying
a symmetry between momentum and winding modes.  Therefore, the notion
of locality in the boundary theory becomes confusing.  Furthermore,
because of this T-duality the boundary theory cannot have a unique
energy momentum tensor.  Such arguments were used in \natistr\ to
argue that the theory of NS5-branes (``little string theory'') is
not a local quantum field theory.  Here we see that this is a common
feature of the boundary theories of all backgrounds of the form \strngm\
(with sufficiently large $d$).

As mentioned above, our discussion applies to some string
backgrounds that were studied in the past.
One example is the ``old matrix model'' (for a review see \morgins), 
which has several interpretations.  It describes the
quantum mechanics of $N\times N$
matrices in the limit $N\rightarrow \infty$.  It can also be thought 
of as a theory of two dimensional gravity.  A third
interpretation arises from interpreting the two dimensional gravity
theory as the worldsheet dynamics of a string. This leads to string
theory in a $1+1$ dimensional spacetime with the  
Liouville field $\phi$ playing the role of a space coordinate. 
The dilaton of this spacetime theory is linear in $\phi$.

In modern language we can say that the equivalence of the large
$N$ matrix model to two dimensional gravity coupled to $c=1$
matter (or equivalently $1+1$ dimensional string theory) is
an example of the holography proposed above. The matrix model
gives a holographic description of two dimensional string
theory\foot{This was suggested by T. Banks several years ago.}.
The observables of the theory can be described in terms of the
matrices and we can compute their Green functions.  These are related
to the S-matrix elements of the bulk spacetime theory which can
be computed using standard worldsheet methods (vertex operators).

The equivalence of the matrix model and $1+1$ dimensional string
theory provides a rather simple example of holography. The arguments
above regarding non-locality of the ``boundary'' theory are
inapplicable here because of the low spacetime dimension; indeed the
boundary theory is standard matrix quantum mechanics.  Similarly, this
example is not well suited for studying the relation between a bulk
theory of gravity and a boundary theory without gravity since the
gravitational sector of the bulk theory (which consists of certain
``discrete states'') essentially decouples from the dynamics.

A richer set of holographic theories of the sort discussed here was
constructed in \ks.  In the notations of \strngm, the compact manifold
$\MM$ in the specific case discussed in \ks\ is a circle\foot{There
are many possible generalizations of the models of \ks\ for which the
manifold $\MM$ is different.}  and the dimension of the boundary $d$
can take the values $d=2,4,6$. The strong coupling singularity at
$\phi\to-\infty$ is removed as in Liouville theory by turning on a
worldsheet superpotential. Consider for example the theory with
$d=4$. It is invariant under eight supercharges which anticommute to
translations in $x$, but not in $\phi$ (of course, translations in
$\phi$ are not a symmetry) or along the circle. It is natural to
conjecture that the full string theory which is naively six
dimensional in this case is equivalent to a four dimensional off-shell
theory without gravity with $\NN=2$ SUSY.

A possible candidate for the ``boundary'' theory is the decoupled
theory of an NS5-brane with worldvolume $\IR^{3,1}\times \Sigma$ with
$\Sigma$ a Riemann surface that can be obtained as follows.  Start
with a configuration of $N$ D4-branes suspended between two parallel
NS5-branes in type IIA string theory (see 
\ref\gk{A. Giveon and D. Kutasov, ``Brane Dynamics and Gauge Theory'', 
hep-th/9802067.}
for a review of the physics of such configurations). 
We can for example take all the
branes to be infinite in $(x^0, x^1, x^2, x^3)$; the fivebranes are
further extended in $(x^4, x^5)$ and the fourbranes are suspended
between them along the $x^6$ axis.  This configuration preserves eight
supercharges and describes at low energies a four dimensional $\NN=2$
SYM theory with gauge group $SU(N)$. Taking the IIA string coupling to
infinity, the above brane configuration turns into an M5-brane in
eleven dimensions with worldvolume $\IR^{3,1}\times\Sigma$ where
$\Sigma$ has been determined in \wttnmqcd. If we now compactify one of
the coordinates $x^7$, $x^8$, $x^9$ on a small circle we get a similar
configuration in IIA string theory. The resulting theory on the
NS5-brane with worldvolume $\IR^{3,1}\times\Sigma$ is not a local 
QFT\foot{A similar construction is described in \wittenmqcdone.},
however it reduces in the infrared to $\NN=2$ SYM with gauge group
$SU(N)$.  

This theory has a few features in common with the $d=4$ theory
constructed in \ks.
The global symmetries of both are naively $U(1)\times U(1)$.  In the
theory of the NS5-brane one $U(1)$ corresponds to rotations in the
$(x^4, x^5)$ plane while the other is the $SO(2)$ subgroup of
rotations of $(x^7, x^8, x^9)$ unbroken by the construction above. 
The second $SO(2)$ symmetry is enhanced to $SO(3)$ in the extreme
infrared limit of the theory on the fivebrane. In the construction of
\ks\ the $U(1)\times U(1)$ symmetry corresponds to momentum and winding
on the circle. One of the $U(1)$ factors is actually broken in both
theories. In the theory of the fivebrane it is broken by quantum
effects (\ie\ at finite QCD scale $\Lambda$).  In the theory of
\ks\ it is broken by the worldsheet superpotential that stops the
theory from running to strong coupling.  Thus we are led to identify
the worldsheet cosmological constant with the QCD scale,
$\Lambda$. The other $U(1)$ remains unbroken in both theories (and as
mentioned above should be enhanced to $SU(2)$ in the extreme IR). 

Another possibility for regulating the strong coupling singularity
at $\phi\to-\infty$ in the theories of \ks\
is to replace the cylinder $\IR\times S^1$
with a Liouville-like superpotential by
the supersymmetric $SL(2)/U(1)$ coset, which changes the topology
to that of a cigar and removes the strong coupling region. 
The symmetry structure of the resulting string theory is the same
as that discussed above.

In the remaining sections we will focus on the specific example of 
string theory in the near-horizon geometry of parallel NS5-branes 
in type II string theory\foot{Our analysis can be easily 
generalized also to other six dimensional theories, such as
the heterotic 5-brane theory.}, 
that was studied by Callan, Harvey and Strominger (CHS) \chs.
We will argue that the theory with vanishing asymptotic string coupling
is dual to the non-local six dimensional
theory without gravity that governs the dynamics of NS5-branes
at vanishing string coupling \natistr. Section 2 contains
rudiments of the relevant classical supergravity solution.
Section 3 contains an analysis of some dynamical issues
in this background. In particular we identify the set of short
representations of the NS5-brane theory with vertex
operators in the weakly coupled string theory regime (the tube of the
CHS theory).  These are given by primaries of the affine $SU(2)$ on
the string worldsheet.  It is known that the NS5-brane theories have 
an $A-D-E$ classification. Our analysis extends the
explanation of \diaco\  
of the $A-D-E$ classification of affine $SU(2)$ modular
invariants.

\newsec{The Near-Horizon Limit of NS5-branes in String Theory}

The decoupled theories on NS5-branes in type II string theory were first
discussed in \natistr, motivated by the study of
compactifications of Matrix theory on high dimensional tori \brs. In
\natistr\ it was argued that the theory on $N$ NS5-branes\foot{We
take $N \geq 2$ for reasons that will be clarified in the next
section.} decouples 
from the bulk in the limit 
\eqn\decouplim{g_s\rightarrow 0,\ \ M_s={\rm fixed},}
because the effective coupling on the NS5-brane is $M_s$, while the
coupling to bulk modes behaves as $g_s$. A DLCQ description of this
theory further supported the existence of a consistent theory that is
decoupled from the bulk \refs{\oens,\edhiggs}.

The existence of a decoupled theory of NS5-branes 
seemed to be in conflict with previous analyses of this
system \refs{\chs,\cghs}. 
In particular, it was pointed out \andyjuan\ that for
an energy density that is finite in string units there is
finite Hawking radiation to the CHS tube region of the 5-brane solution,
suggesting that the theory does not decouple from the fields
in this region.

The proposal in \juan\ can be used to reconcile the two points of
view\foot{Similar ideas have also been suggested by various people,
including C. V. Johnson,
J. Maldacena and A. Strominger.}. According to this
conjecture a $d$ dimensional theory without gravity, such as our
theory for $d=6$, may be dual to a higher dimensional theory with
gravity. In our case, string or M-theory in the background 
of the 5-brane solution in the limit 
\decouplim, including the CHS tube, is
conjectured to be dual to the decoupled theory on the NS5-branes. In
particular, the fields in the tube which arise in the Hawking
radiation \andyjuan\ are interpreted as part of this decoupled
theory. Correlation functions of observables in this theory may be
defined by setting appropriate boundary conditions at the weak
coupling boundary of the CHS background.

In this section we will discuss the classical supergravity 
solution which
arises in the limit \decouplim\ of the metric of NS5-branes.
M theory in this background
is conjectured to be dual to the
decoupled NS5-brane theory. Parts of this section overlap with
\refs{\sunny,\townsend}.

\subsec{IIA and M Theory 5-branes}

We start by discussing NS5-branes of type IIA string theory, which
may be viewed as M-theory 5-branes localized on the eleven dimensional
circle. Thus, they are described by the metric for $N$ 
M5-branes at a point on a transverse circle. M-theory has a scale
$l_p$, and the radius of the circle asymptotically far away from the
5-branes will be denoted by $R_{11}$ (we
will not be careful about numerical factors). The (asymptotic) string
scale is given by $R_{11} l_s^2 = l_p^3$.  We are interested in taking
the limit \decouplim\ in which $l_p$ and $R_{11}$ go to zero with
$l_s$ kept fixed. $l_s$ will then be the dimensionful parameter of
the decoupled theory on the NS5-branes \natistr. 
The near-horizon metric for $N$ overlapping
5-branes in this configuration is given by :
\eqn\mmetric{ds^2 = H^{-1/3} [ dx_6^2 + H (dx_{11}^2 + dr^2 + 
r^2 d\Omega_3^2) ],}
where
\eqn\hdef{H = \sum_{n=-\infty}^{\infty} { N l_p^3 \over 
{(r^2 + (x_{11} - n R_{11})^2)^{3/2}}},} 
$x_{11}$ is periodic with
period $R_{11}$, and $dx_6^2$ is the metric on $\IR^{5,1}$. This is the
metric for $N$ overlapping 5-branes -- the generalization to
5-branes located at different $x_{11}$ positions is straightforward.
The supergravity solution involves also a 4-form field strength with
$N$ units of flux, which we will not write explicitly.

In the limit \decouplim\ the natural coordinates to define are such 
that the tension of a string arising from an M-theory membrane
stretched between 5-branes at distances $r$ or $x_{11}$ remains
constant. Thus, we
choose $U = r / l_p^3$ and $y_{11} = x_{11} / l_p^3$. Both of these
coordinates have dimensions of mass squared, and the coordinate
$y_{11}$ has a periodicity of $R_{11} / l_p^3 = 1 / l_s^2$. In terms
of these variables the metric is:

\eqn\newmetric{ds^2 = l_p^2 \tH^{-1/3} [ dx_6^2 + \tH (dy_{11}^2 + dU^2 +
U^2 d\Omega_3^2) ]}
with
\eqn\newh{\tH = \sum_{n=-\infty}^{\infty} {N \over {(U^2 + (y_{11} - 
n/l_s^2)^2)^{3/2}}}.}  Except for an overall factor of $l_p^2$ in
front of the metric, it remains finite in this limit. As in the case
of a similar factor in the $AdS_5\times S^5$ metric \juan, this $l_p$
will drop out of any physical computations. Below we will study the
conjecture that M-theory on the manifold \newmetric\ is equivalent
(``dual'') to the six dimensional theory of $N$ NS5-branes in type IIA
string theory. Both theories have $(2,0)$ six dimensional SUSY (four
supercharges in the ${\bf 4}$ of $Spin(5,1)$) and a global $SO(4)$
R-symmetry.

There are two regions where the metric \newmetric\ simplifies
considerably. It is known that 
distances that are large compared to $\sqrt{N} l_s$ in the six
dimensional theory of NS5-branes
correspond to the $(2,0)$ SCFT, the extreme IR limit of the NS5-brane
theory. Equivalently, we can approach this low-energy limit by 
taking $l_s\to
0$, which corresponds in
\newmetric\ to small values of $U$ and $y_{11}$ (compared to
$1/l_s^2$). In this limit the sum in
\newh\ is dominated by the contribution from $n=0$, and the metric
\newmetric\ becomes the metric for $AdS_7\times S^4$, as in \juan, 
which is indeed believed to be dual to the $(2,0)$ SCFT. The six
dimensional Poincare symmetry is enhanced to the conformal group, and
the $SO(4)$ global R-symmetry is enhanced to $SO(5)$.

The second interesting limit is large $U$.
For $U \gg 1/l_s^2$, the sum in \newh\ can be approximated by an integral,
and the result is
\eqn\farmetric{ds^2 = l_p^2 {U^{2/3} \over {(N l_s^2)^{1/3}}}
[ dx_6^2 + {N l_s^2 \over U^2} (dU^2 + dy_{11}^2) + N l_s^2
d\Omega_3^2].}  
For $U\gg \sqrt{N}/ l_s^2$ the theory becomes
weakly coupled type IIA string theory. The quantity in the square
parentheses (without the $dy_{11}^2$ term) is exactly the type IIA
string metric, and the IIA string coupling is
\eqn\gs{g_s^2(U) = {N \over l_s^4 U^2}.}

Furthermore, for large $N$ the curvatures are small (either in the
eleven dimensional metric or in the ten dimensional metric) for any
value of $U$, so we can use the low-energy supergravity to compute
some properties of the type IIA NS5-brane theories.

We will sometimes find it useful to work, in the weakly coupled string
theory regime, with a new coordinate $\phi$ which is
\eqn\defphi{U l_s^2 / \sqrt{N} = e^{\phi / \sqrt{N} l_s}.}
This brings the metric to the more familiar linear dilaton form 
\eqn\phimetric{ds_{string}^2 = dx_6^2 + d\phi^2 + N l_s^2 d\Omega_3^2,\ \ 
g_s^2(\phi) = e^{-2\phi/\sqrt{N} l_s}.}

\subsec{Energy Scales in the Theory of Type IIA NS5-branes}

There are several energy scales that could be important in the
discussion of the decoupled theory of type IIA NS5-branes.  The first
scale is $1/\sqrt{N} l_s$, which is the scale that appears explicitly
in the metric in the linear dilaton region. This scale also appears in
previous computations of various properties of the NS5-brane theories;
for instance, their Hagedorn temperature is $T = 1/ \sqrt{N} l_s$. 

We can find additional energy scales in the problem by examining
cross-over regions as we change $U$. Naively we would interpret a position
in the $U$ coordinate as corresponding to an energy scale $U \sim
E^2$. However, because there is another dimensionful parameter in our
problem, $l_s$,
we can also relate effects happening at some position $U$ to
physical processes at energies $E \sim \sqrt{U} f(Ul_s^2)$ for some
function  $f$. To be precise one must
define the process of interest first, and the scale $U$
may in principle appear in different ways in different processes.

As we change $U$, we can identify the following cross-over scales
where the behavior of the theory changes :
\item{(1)} One special point is where $g_s=1\Rightarrow 
U\sim\sqrt{N} / l_s^2$.  At this scale we go over from weakly coupled
type IIA string theory to a strongly coupled theory (\ie\ M-theory). 
\item{(2)} Another scale 
is the place where the radius of the $y_{11}$ circle is of the same
order as the radius of the $S^3$, naively indicating a crossover to an
$AdS_7\times S^4$ regime of the theory. In the metric written above
this happens at $U \sim 1/l_s^2$. At this scale $U$ is of the same
order as the periodicity in $y_{11}$, and we can no longer approximate
the sum over $n$ in \newh\ by an integral.

\subsec{The Type IIB NS5-brane}

The behavior of type IIB NS5-branes in the limit \decouplim\ is
similar to the IIA solution in the linear dilaton regime, but it is
very different close to the 5-branes, as in the limit \decouplim\ the
IIB solution becomes singular close to the 5-branes.

The behavior of the metric in different regimes is analyzed in \sunny.
The string metric far from the branes is the same as the one in
\farmetric, but now it is more natural to define the $U$ coordinate
as $U = r / g_s l_s^2$ which is the mass of a D-string stretched
between two NS5-branes, and from it define $\phi$ as in \defphi. In
terms of this coordinate the string metric and coupling are as in
\phimetric.
As we decrease $U$, we encounter the first crossover scale at $U
\sim \sqrt{N} / l_s$, where $g_s \sim 1$. At this scale the string
coupling becomes large, but the curvatures (in the Einstein metric)
are still small, so we can go over to an S-dual picture \sunny. 
In the dual
picture the string metric is the same as above, multiplied by $1/g_s
\sim U$, and the coupling behaves as $\tilde{g}_s^2 = 1/g_s^2 \sim
l_s^2 U^2 / N$. In this new description, the string coupling becomes
smaller and smaller as we decrease $U$, but the curvature (for example
of the $S^3$) becomes larger and larger. The curvature becomes
Planckian at the scale $U \sim 1 / \sqrt{N} l_s$, so beyond this scale
we can no longer trust supergravity. 
This agrees with our expectations, since
the low-energy gauge coupling is given by $g_{YM}^2 = l_s^2$, so we
expect perturbation theory to be valid whenever the dimensionless
coupling $g_{YM}^2 N U = N
l_s^2 U$ is small (interpreting $U$ as an energy scale in the SYM
theory). Thus, we can interpret this scale as corresponding to the
breakdown of the SYM perturbation theory \sunny.

\newsec{Observables and Correlation Functions of the NS5-brane Theories}

The six dimensional NS5-brane theory has a scale, $l_s$. In the dual
description this scale appears in the metric, \eg\ as in \newmetric,
\newh\ for the IIA case. As discussed above, to study the long
distance behavior of the theory we have to analyze it in the limit
$l_s\to 0$, where it is governed by a local QFT, the $(2,0)$ SCFT for
IIA fivebranes and the IR free SYM with sixteen supercharges for
IIB. If the fivebrane theory had been a local QFT for all energy
scales, the short distance behavior would have been governed by a UV
fixed point. This fixed point would have been studied by taking
$l_s\to\infty$ in \newmetric, \newh. In our case this limit leads to
the linear dilaton geometry \phimetric.

Because string theory in the linear dilaton geometry \phimetric\ cannot 
describe a field theoretic
UV fixed point, the NS5-brane theory is not a local QFT. 
However, we do expect the fivebrane theory to have the
property that, as in local QFT, observables are defined in the UV
region \phimetric.  Therefore, as a check of the duality, we next
discuss the spectrum of excitations of string theory in the linear
dilaton vacuum and compare it to the set of observables of the
NS5-brane theory. We mainly focus on short representations of
supersymmetry, since the complete list of those is independently known
in the fivebrane theories.

To find these short representations, recall that 
``little string theories'' with sixteen supercharges
have an $A-D-E$ classification. In the usual description the 
different theories may be obtained by studying decoupling limits 
of type II string theory on $K3$ with $A-D-E$ type singularities. 
In the CHS limit \phimetric\ they correspond to the $A-D-E$
classification of modular invariants of $SU(2)$ WZW models \diaco.

The global symmetry of 
these theories is $SO(4)$. The moduli spaces of vacua are $(S^1 \times 
\IR^4)^r / \WW$ for the type IIA theory and $(\IR^4)^r /
\WW$ for the type IIB theory, where $r$ is the rank of the $A-D-E$ group
and $\WW$ is its Weyl group.  The global $SO(4)$ symmetry acts on the
$\IR^4$ factors. This suggests that the special chiral representations
are correlated with $\WW$-invariant products of scalars, which are
the natural coordinates on the moduli space.  In the type IIB
theory they are easily identified as the gauge invariant polynomials
in the scalars in the gauge multiplets; \ie\ they are identified with
the Casimirs of the gauge group.  Their $SO(4)$ representation is a 
traceless symmetric tensor whose order is the order of the corresponding
Casimir. 

A similar result is expected in the IIA theory. This can
be shown by compactifying it to five dimensions, where the low energy
theory is an $A-D-E$ gauge theory. Therefore, the representations are
traceless symmetric $SO(5)$ tensors whose orders are those of the
Casimirs. For the $A_n$ theories the same conclusion can be reached by 
using their DLCQ\foot{The extension to the $D_n$, $E_n$
theories leads to predictions about the cohomology with compact
support of the moduli spaces of $D_n$ and $E_n$ instantons on
$\IR^4$.} formulation \abs. For the 
$A$, $D$ models with large $N$ one can also use the M-theory
duals of the relevant SCFTs \refs{\aoy\lr\minwalla-\halyo}. 
Of course, only an $SO(4)$ out of the $SO(5)$ R-symmetry of the 
$(2,0)$ SCFT is visible at large $U$.

In the next subsection we discuss string theory
in the linear dilaton CHS background \phimetric\ and show that the
spectrum of short representations of supersymmetry is identical
to that described above.

\subsec{Worldsheet aspects of six dimensional string theory}

The CHS background \phimetric\ is:
\eqn\aa{\IR^{5,1}\times \IR\times S^3_N.}
$\IR^{5,1}$ is the six dimensional spacetime of the
$(2,0)$ theory; the remaining four dimensions
parametrize the space transverse to the fivebranes. 
The second factor in \aa\ is the radial direction
$\phi$ or $U$ \defphi. It is described on the worldsheet
by a free field whose stress tensor has an improvement
term,
\eqn\bb{T_\phi=-{1\over2}(\partial\phi)^2-{Q\over2}
\partial^2\phi;\;\;\;Q=\sqrt{2\over N}.} 
The three-sphere $S^3_N$ of radius $\sqrt{N} l_s$ describing the
angular coordinates corresponds to an $SU(2)$ WZW model with
level $k=N-2$ (note that this requires $N \geq 2$). 

In addition to the above bosonic worldsheet fields, the
theory also has free worldsheet fermions. 
The worldsheet fields $X^\mu(z,\bar z)$
($\mu=0,1,2,\cdots, 5$) parametrizing the six dimensional
spacetime are accompanied by left and right moving superpartners
$\chi^\mu(z)$, $\bar\chi^\mu(\bar z)$; the worldsheet superpartners
of $\phi$ and the $SU(2)$ WZW are $\psi^0(z)$, $\bar\psi^0(\bar z)$ 
and $\psi^i(z)$, $\bar\psi^i(\bar z)$ ($i=1,2,3$), respectively.

The $SO(4)\simeq SU(2)_R\times SU(2)_L$ isometry of the
three-sphere in \aa\ acts on the worldsheet fields as follows.
The bosonic $SU(2)_k$ WZW model contains (anti-) holomorphic
currents $J^i_B(z)$, $\bar J^i_B(\bar z)$ which generate
an $SU(2)\times SU(2)$ symmetry. 
The fermions $\psi^i(z)$, $\bar\psi^i(\bar z)$ transform 
in the adjoint of an $SU(2)\times SU(2)$ symmetry, generated
by the level two currents 
\eqn\cc{J^i_F={1\over2}\epsilon^{ijk}\psi_j\psi_k;\;\;\;
\bar J^i_F={1\over2}\epsilon^{ijk}\bar\psi_j\bar\psi_k.}       
The total currents
\eqn\dd{J^i=J^i_B+J^i_F;\;\;\;\bar J^i=\bar J^i_B+\bar J^i_F}
generate the above $SO(4)$ symmetry. The total level of the
currents $J^i$, $\bar J^i$ is $(N-2)+2=N$. 

The spacetime supercharges of the CHS near-horizon string theory
are a subset of those of the full IIA string theory.
Denoting by $\alpha\in{\bf 4}, \bar\alpha\in {\bf\bar 4}$
and $a\in{\bf 2}, \bar a\in {\bf\bar 2}$ spinor indices
of $SO(5,1)$ and $SO(4)$, respectively, the thirty two 
supercharges of IIA string theory transform as $Q_{\alpha a}$,
$Q_{\bar\alpha \bar a}$, $\bar Q_{\alpha\bar a}$, 
$\bar Q_{\bar\alpha a}$ ($Q$, $\bar Q$ arise from
left, right movers on the worldsheet, respectively). By using the form
of the gauged worldsheet $\NN=1$ superconformal generators:
\eqn\ee{\eqalign{
T=&-{1\over2}(\partial X^\mu)^2-{1\over2}\psi^\mu\partial\psi^\mu
+T_\phi-{1\over N}J^i J^i-{1\over 2}\partial\psi^a\psi^a\cr 
G=&\psi^\mu\partial X^\mu+\psi^0\partial\phi+
\sqrt{2\over N}(\psi^i J^i+\psi^1\psi^2\psi^3-\partial\psi^0)\cr
}}
and the similar formulae for the other worldsheet chirality,
one can show that the physical supercharges in the background
\aa\ are $Q_{\alpha a}$, $\bar Q_{\alpha \bar a}$, which generate
$(2,0)$ six dimensional SUSY. 

Note that we could have studied the CHS limit of NS5-branes 
in type IIB string theory as well. In that case the surviving
SUSY would have been $(1,1)$ and we would have obtained the other
six dimensional string theory discussed in \natistr. The low energy
limits of the $(2,0)$ and $(1,1)$ string theories are very different.
While the former flows in the infrared to the non-trivial $(2,0)$
field theory, the latter reduces at low energies to the (infrared
free) six dimensional SYM. From the point of view of the
string theory in the tube limit \aa, this difference has to do with
the different ways the IIA and IIB theories
treat the strong coupling region at small $\phi$.

The following general features are clear from the above description. 

\item{(1)} The NS5-brane theory has a stress
tensor. In the (dual) string theory of the CHS tube this is the statement
that the theory has six dimensional gravitons (generalizing the
identification of the graviton with the energy-momentum
tensor in the AdS/CFT correspondence \refs{\gkp,\wttn}).
\item{(2)} Upon compactification on tori, the NS5-brane
theory has T-duality \natistr. In our description this arises from
the T-duality of the dual IIA or IIB string theory. 
Thus, it cannot be a local QFT. In particular, upon compactification the
identification of the graviton is not unique (since it varies when we
T-dualize), corresponding to the non-uniqueness of the energy momentum
tensor in the six dimensional theory \natistr.

We next turn to the spectrum of physical operators of the string
theory in the CHS tube.  We will focus on short representations of the
relevant SUSY algebra. As a first example consider the $A$ series of
six dimensional string theories, which is conjectured to be dual to
the theory with the $A$ modular invariant of $SU(2)$ in the tube
limit.  The only primaries of $SU(2)_R\times SU(2)_L$ that appear in
the $A$ modular invariant of $SU(2)_k$ are $V_{j,j}$ with spin $j$ for
both $SU(2)$'s, with $2j=0,1,2,\cdots, k$ (recall that the level $k$
is related to the number of fivebranes $N$ via $k=N-2$).  Each such
primary gives rise to a short representation.  As is clear from our
previous comments on the global symmetries, the state with spin $j$
transforms as a traceless symmetric tensor with $2j$ indices under
$SO(4)$.

To obtain the bounds on $j$ we need to describe the
physical states slightly more precisely. The lowest component
of each representation is a scalar under six dimensional
Lorentz. The corresponding vertex operator takes the form
(in the $-1$ picture)
\eqn\verop{\psi\bar\psi V_{jj}
e^{\beta_j\phi}}
The fermions $\psi$, $\bar\psi$ transform
as the $\bf(3,1)$ and $\bf(1,3)$ of 
$SU(2)_L\times SU(2)_R$; $\beta_j=j\sqrt{2\over N}$. 
The indices under each $SU(2)$ are contracted between
the fermions and $V_{jj}$ to form representations
with spin $j-1$, $j$ and $j+1$. One can show that the
representation with spin $j$ is unphysical, while that
with spin $j+1$ gives rise to the lowest component of
a short representation. The states with spin $j-1$ are
descendants in this representation.
Thus, one finds short multiplets in the $SU(2)_R\times SU(2)_L$
representations of spin
$(j+1, j+1)$ with $2j=0,1,2,\cdots, N-2$ as above.
In $SO(4)$ language these are traceless symmetric tensors with
$2j+2=2,3,\cdots, N$ indices. This agrees with our general
expectation in terms of the Casimirs of $A_{N-1}$. Note that in
our description we can explicitly see the truncation of the chiral
operators at $2j+2=N$, which is generally obscure in the AdS/CFT
correspondence. This provides further evidence for the validity of
the bulk/boundary correspondence for finite values of $N$.

Returning to the $SO(4)$ vs.\ $SO(5)$ issue, we expect (just like
in type IIA string theory in flat spacetime) the symmetric
tensors of $SO(4)$ to be extended to symmetric tensors of $SO(5)$
with non-perturbative states in the CHS string theory, \ie\ 
D-branes. The missing states are exactly the D0-brane states, which
indeed exist only for type IIA. Adding these states we find, for large
$N$, the full spectrum of the low-energy SCFT. We have not shown 
that for finite $N$ the spectrum of states involving D0-branes
truncates at the appropriate value of $j$.

The D-series models exhibit some new features.
Recall that D-series modular invariants in $SU(2)$ WZW CFT
exist only for even $k$ and they have the following spectrum
of primary operators: 
\item{(1)} Operators $V_{j,j}$ with equal left and right spins
which are both {\it integer} (and as usual bounded from above
by $k/2$). 
\item{(2)} A single additional operator $V_{j,j}$ with $j=k/4$.
\item{(3)} A series of operators $V_{j_1, j_2}$ with 
$j_1={k\over4}+n$ and $j_2={k\over 4}-n$ with integer $n$.

The corresponding low energy theory is an $SO(2N)$ gauge theory with
sixteen supercharges for type IIB, and the $D_N$ $(2,0)$ SCFT (which
is dual to M-theory on $AdS_7\times RP^4$ \aoy) for type IIA 
(the relation between $k$ and $N$ is in this case $k=2(N-2)$). The
first kind of operators in the tube string theory above are
interpreted as before in the two low energy theories.  The operator
(2) above is also easy to interpret: it corresponds to the Pfaffian
operator in the $SO(2N)$ gauge theory that appears for type IIB, and
to a similar operator of dimension $\Delta = 2N$ in the $(2,0)$
SCFT for type IIA. In M-theory on $AdS_7\times RP^4$
this operator corresponds to an $M2$-brane wrapped
around an $RP^2$ in $RP^4$ (as in \edbaryon).

The operators (3) are interesting. Since their worldsheet
left and right scaling dimensions differ by an integer
they are similar to Dabholkar-Harvey \dh\ states in toroidal
compactifications of string theory and, just like the above,
they give rise to a large number of ``medium''
multiplets. It would be interesting to understand whether
supersymmetry is sufficient  
to guarantee their appearance in the low energy theory,
and, if the answer is yes, to identify them in the low energy
IIA and IIB theories.

One can also study the $E_6$, $E_7$ and $E_8$ six dimensional
string theories. These theories have the following structure
in the tube limit. The bosonic WZW model contains operators 
with $SU(2)_R\times SU(2)_L$ spin $(j,j)$ with the following 
values of $2j$:
\eqn\esse{\eqalign{
E_6:\;\;\;& 2j=0,3,4,6,7,10\cr E_7:\;\;\;& 2j=0,4,6,8,10,12,16\cr
E_8:\;\;\;& 2j=0,6,10,12,16,18,22,28.\cr }} Each of these gives rise,
as in the $A$, $D$ series above, to short multiplets in the six
dimensional string theory.  At low energies the spectrum of chiral
primaries should be compared to the six dimensional IR free gauge
theory with sixteen supercharges and $E_n$ gauge group for type IIB,
and to the $E_n$ (2,0) SCFT for IIA. The latter cannot be described by
eleven dimensional supergravity since the curvature of the relevant
eleven manifold is large in Planck units, but our general
considerations above suggest that again the special representations
are traceless symmetric $SO(5)$ tensors with the number of indices
related to the order of the $E_n$ Casimir.  It is easy to see that the
results predicted by \esse\ are indeed correlated with these Casimirs
in the right way.

The analysis of \diaco\ gave an explanation of the origin of the $A-D-E$
classification of affine $SU(2)$ modular invariants.  The results here
explain another aspect of this classification which is not obvious
from purely conformal field theory considerations.  It clarifies why
the spinless primaries are correlated with the Casimirs of $A-D-E$
and, therefore, why their number is given by the rank of the $A-D-E$
group. 

The $E_n$ theories, just like the $D_n$ ones,
have states $V_{j\bar j}$ with
$j\not=\bar j$. Thus the issue of the existence of medium
representations in the infrared theory raised in that context   
must be understood here as well.

\subsec{Branes in the near-horizon geometry}

In addition to the observables described in the previous
subsection, branes can also
propagate in the backgrounds described above. It is interesting to
analyze the brane spectrum in the geometry we find, and to interpret it
in terms of the dual NS5-brane theory.

Let us start with the small $U$ limit of the IIA case, where the
theory goes over to the $AdS_7\times S^4$ compactification of M-theory,
which is conjectured to be dual to the $(2,0)$ 6D SCFT. The
interesting branes in this limit seem to be the membrane and the
M-theory 5-brane wrapped around $S^4$. The membrane tension does not
depend on the $U$ coordinate; the $U$ coordinate is defined so that
the mass of a membrane stretched in the $U$ direction is
linear\foot{Note that the coordinate $U$ we use here is the square of the
coordinate $U$ in \juan.} in $U$. This enables us to look at
configurations of membranes ending on some surface at the boundary of
$AdS$ (large $U$); the energy of these configurations is linearly
divergent in $U$, and they may be identified with ``Wilson surface''
observables of the $(2,0)$ SCFT.

On the other hand, a 5-brane wrapped on $S^4$ cannot exist on its own,
like the wrapped 5-branes in the $AdS_5\times S^5$ background
\edbaryon.  This is because the background 4-form field acts as a source
of 2-form charge in the 5-brane, which must be balanced by $N$ M
theory membranes ending on such a 5-brane in the $AdS_7$ (note that
the wrapped 5-brane is a string in the $AdS_7$). These membranes must
have another boundary, so they must stretch to the boundary of $AdS$
(or to another 5-brane with opposite orientation); thus this
configuration behaves like a baryon which contributes to the product
of $N$ ``Wilson surface'' observables.

Both types of branes described above exist also in the full background
described in the previous section. In the linear dilaton region, the
membrane becomes a D2-brane of type IIA string theory, whose tension
(in 11D Planck units) is still independent of $U$. These
membranes/D2-branes may be used to define ``Wilson surface''
observables of the type IIA NS5-brane theories, by analyzing
configurations with D2-branes stretching to infinity in the $U$
direction. At large distances (measured in string units) these
observables will go over to the ``Wilson surface'' observables of the
low-energy $(2,0)$ SCFT \juanwilson.

Similarly, the 5-brane wrapped on $S^4$ becomes a D4-brane wrapped on
$S^3$. The background NS 3-form field induces a magnetic charge in the
4-brane field theory, which must be balanced by $N$ D2-branes, as
before. So, this state still behaves as a generalized baryon
vertex.

We will not discuss here the most general possible branes, but it is
interesting to analyze D-branes which stretch only in the $\IR^{5,1}$
directions (in the linear dilaton region). The mass of such D-branes
behaves (in string units) like $1/g_s$, so they are not stable objects
but instead tend to fall into the strong coupling region. This
agrees with the expectation \natistr\ that D-branes in $\IR^{5,1}$ will
form bound states with the NS5-branes; we can interpret the branes
falling in to small values of $U$ as a bound state of the D-branes with
the NS5-branes, which tends to spread out in the $\IR^{5,1}$ directions
(using the interpretation of $U$ as an inverse distance scale).

The analysis is similar for the type IIB theory. In this theory the
D1-brane has (by construction) finite tension for large $U$, and can
be used to define Wilson line operators (which go over to Wilson lines
in the 6D SYM theory at large distances). A D3-brane wrapped around
the $S^3$ serves as a baryon vertex, since $N$ D-strings have to end
on such a D3-brane. D-branes in $\IR^{5,1}$ can be interpreted as in the
previous paragraph. For very small values of $U$ it should be possible
to interpret the fundamental string (stretching in the $\IR^{5,1}$
directions) as an instanton of the SYM theory, and the $U$ coordinate
as the scale size of this instanton (this is similar to the
interpretation of D-instantons in the $AdS_5\times S^5$ background
\refs{\gb\bgkr-\bklt}), 
but it is not clear if such an interpretation makes sense
also for large $U$. 

\subsec{Holographic Computation of a Scalar Correlator in the 
IIA NS5-brane
Theory}

In this subsection we will discuss some aspects of the Euclidean
2-point function of a scalar field in the theory of the IIA
NS5-branes. We will do the computation in low-energy supergravity,
implying the following limitations:

\noindent 1. The momentum has to be well below the string scale 
$p\ll 1/l_s$. Otherwise there could be large corrections to the
supergravity analysis.

\noindent 2. We require that $N$ be large so that the various 
curvatures in the solution are small.

Even under these restrictions there are two distinct regimes of
momenta, compared to the scale $1/\sqrt{N}l_s$. At low momenta
compared to this scale we will reproduce the results of the low-energy
$(2,0)$ SCFT, whereas for $p>1/ \sqrt{N}l_s$ (but still much smaller
then $1/ l_s$) we will start seeing the effects of the ``little
string theory''.

Note that we do not really have a scalar field in the problem (which
is a scalar in 11D), but this simpler computation should have the same
qualitative features as the correlation functions of other fields in
the theory (such as the graviton, which is related to the six
dimensional energy momentum tensor). 

We will follow the procedure of \gkp\ for computing 2-point functions
of scalar fields. We should start by finding the solution to the equation
of motion for a scalar field in the background \newmetric. Assuming
no dependence on the $S^3$ coordinates, and an $e^{ip\cdot x}$ dependence
on the $\IR^{5,1}$ coordinates, the equation of motion is
\eqn\eom{[\del_U(U^3 \del_U) + U^3 \del_{11}^2 - {\tilde H} U^3 p^2] 
\Phi(U,
y_{11}) = 0,}
where $\tilde H$ was defined in \newh.

We have not been able to solve \eom\ exactly, but dimensional analysis
of this equation reveals much of its physical content. The factor of
$N$ appears only in the last term, hence the dependence of the
solution on $p$ and on $N$ will be via the combination
$p^2N$. Furthermore, if we rescale $U$ and $y_{11}$ by $l_s^2$, making
them dimensionless, then the differential operator in \eom\ depends
only on $Nl_s^2p^2$. This is true in all the regions of the SUGRA
background. If we are interested in momenta below $1/ \sqrt{N}
l_s$, then we may hold $p$ fixed and take $l_s\rightarrow 0$, in
which case the differential equation becomes that of a scalar
field on $AdS_7\times S^4$ and we will reproduce the results of the
$(2,0)$ SCFT. When $l_s$ is not strictly zero, the correlator
will deviate from that of the SCFT, and this deviation will be governed
in momentum space by powers of $Nl_s^2p^2$. Thus, eleven dimensional
supergravity can be used to analyze the behavior of the NS5-brane theory
away from the IR fixed point, for $1/\sqrt{N} l_s < p  \ll 1/l_s$.

In the low
momentum regime $p\ll 1/ \sqrt{N} l_s$ we can make the
computation more explicit. This will also serve to
show that the dependence on the UV cutoff (which one takes to infinity
only at the end of the computation of a correlator) does not
invalidate the conclusion of the previous paragraph. In the large $U$
region the radius of the $y_{11}$ direction vanishes and therefore
$\partial_{11}
\Phi =0$.  Thus, we get the equation
\eqn\mtna{[\del_U(U^3 \del_U) - N l_s^2 p^2 U] \Phi(U) = 0,}
which has two independent solutions,
$\Phi_\pm \sim U^{\beta_\pm}$, where 
$\beta_\pm = -1\pm\sqrt{1+ N l_s^2 p^2}$.

In the small $U,y_{11}$ region (the $(2,0)$ SCFT region) the metric looks 
like $AdS_7\times S^4$, so we have the standard solution for a scalar
field in $AdS_7$. The solution on this space which is regular at $U=0$ is
\eqn\adssol{\Phi_0(U,y_{11}) = (\sqrt{Np^2} z)^3 K_{3}(\sqrt{Np^2}z),}
where $z = (U^2 + y_{11}^2)^{-1/4}$ and $K_3$ is a Bessel function.
In particular, the solution in this regime has an asymptotic expansion
in $\sqrt{Np^2} z$, of the form
\eqn\adssolapp{\Phi_0(z) \sim 1 + a_1 N p^2 z^2 + a_2 (N p^2 z^2)^2 +
a_3 (N p^2 z^2)^3 (log(N p^2 z^2) + a_4) + \cdots} for some constants
$a_i$. This asymptotic expansion can be used when both $\sqrt{Np^2} z
\ll 1$ and $U \ll 1/l_s^2$, which implies that $p^2 \ll 1 / N l_s^2$,
so we are at very low momenta where we expect the result to be similar
to the result for the $(2,0)$ SCFT.

As in \gkp, we will set a cutoff $U_0$ for the theory, and compute the
2-point function with this cutoff, taking $U_0 \to \infty$ at the end.
For $U_0 \gg 1/l_s^2$, the solution near $U_0$ will be some linear
combination of the solutions $\Phi_\pm$
described above, which is determined by
the fact that it should go over to $\Phi_0$ for small $U$. Thus, the
solution for large $U$ will be of the form
\eqn\solution{\Phi(U) \sim {{a_+(p) U^{\beta_+} + a_-(p) U^{\beta_-}} 
\over 
{a_+(p) U_0^{\beta_+} + a_-(p) U_0^{\beta_-}}},}
where $a_\pm(p)$ are some functions which are determined by the form
of the exact solution, and we normalized the solution so that 
$\Phi(U_0) = 1$ (as in \gkp).

As in \gkp, we would now like to evaluate the action for the scalar
field.  It reduces to a boundary term at $U=U_0$, which is of the form
$F(p) \propto U^3 \Phi \del_U \Phi$ (evaluated at
$U=U_0$). Substituting 
in the solution \solution, we find that for large $U_0$ this behaves
like
\eqn\twopoint{F(p) \propto U_0^2 (\beta_+ + (\beta_- - \beta_+)
{a_-(p) \over a_+(p)} U_0^{\beta_- - \beta_+} + \cdots).}
The two point function in position space will be given by the Fourier
transform of the leading non-analytic term in $F(p)$ (in the $U_0 \to
\infty$ limit). The expansion \adssolapp\ suggests that we can expand
the whole solution in an asymptotic expansion in $p^2$, for which the
leading non-analytic term would be at the order $p^6 \log(p^2)$, with
additional non-analytic terms of the form $p^{6} (p^2 N l_s^2)^n
\log(p^2)^k$ (times some function of $U$). This was shown to be true
for the solution in the full 3-brane metric which interpolates between
$AdS_5\times S^5$ and flat Minkowski space in \gh, and we expect a
similar behavior here.
Thus, the function
$a_-(p)/a_+(p)$ should have a similar asymptotic expansion, so that
the leading non-analytic term in \twopoint\ is of the order of $p^6
\log(p^2)$ (note that this leading term is independent of $U_0$),
leading to a 2-point function which behaves like $1 / |x - y|^{12}$.
This is exactly the behavior we expect for a dimension 6 operator
(corresponding to a massless scalar field in supergravity) in the
low-energy SCFT. The procedure above enables us to compute also the
first corrections to this low-energy expression, and we immediately
see that they will depend on $Nl_s^2/|x-y|^2 $, as expected. 

\bigskip
\centerline{\bf Acknowledgements}

We would like to thank A. Giveon, J. Maldacena, E. Martinec,
S. Shenker and A. Strominger for useful discussions. O.A. also thanks
the Aspen Center for Physics for hospitality during the final stages
of this work.  The work of O.A. was supported in part by
\#DE-FG02-96ER40559, that of M.B.  by NSF-PHY-9513835, that of D.K.
by \#DE-FG02-90ER40560 and that of N.S.  by grant \#DE-FG02-90ER40542.

\listrefs

\end